\documentclass[amsmath,amssymb,pre,twocolumn,a4paper]{revtex4}
\usepackage{amsfonts} 
\usepackage{amsmath}
\usepackage{graphicx} 
\usepackage{overpic} 
\usepackage{color}

\usepackage{xr}
\newcommand{\be}{\begin{eqnarray}} 
\newcommand{\ee}{\end{eqnarray}}
\newcommand{\D}{\mathrm{d}}

\begin{document}

\title{Optimal random deposition of interacting particles}

\author{Adrian Baule$^1$}

\affiliation{
$^1$School of Mathematical Sciences, Queen Mary University of London,  London E1 4NS, UK
}
\email{a.baule@qmul.ac.uk}

\begin{abstract}

Irreversible random sequential deposition of interacting particles is widely used to model aggregation phenomena in physical, chemical,  and biophysical systems. We show that in one dimension the exact time dependent solution of such processes can be found for arbitrary interaction potentials with finite range. The exact solution allows to rigorously prove characteristic features of the deposition kinetics, which have previously only been accessible by simulations. We show in particular that a unique interaction potential exists that leads to a maximally dense line coverage for a given interaction range. Remarkably, this distribution is singular and can only be expressed as a mathematical limit. The relevance of these results for models of nucleosome packing on DNA is discussed. The results highlight how the generation of an optimally dense packing requires a highly coordinated packing dynamics, which can be effectively tuned by the interaction potential even in the presence of intrinsic randomness.

\end{abstract}

\maketitle

The deposition of particles on a substrate is a ubiquitous phenomenon in science and engineering \cite{Elimelech:1995aa}. From a theoretical perspective, the deposition dynamics is widely modelled in terms of a random sequential adsorption (RSA) process, which represents a paradigmatic adsorption mechanism \cite{Evans:1993ab}. Since Renyi's seminal work on the ``car parking problem" (the RSA of equal line segments in 1$d$) \cite{Renyi:1958aa,Renyi:1963aa}, RSA models and their variants have been successfully used to model polymer and colloid adsorption on surfaces \cite{Feder:1980aa,Talbot:2000aa,Cadilhe:2007aa} and are also relevant in other contexts such as the compactification of granular matter \cite{Nowak:1998aa,Tarjus:2004aa}, genome sequencing \cite{Roach:2000aa}, and nucleosome packing on DNA \cite{Kornberg:1988aa,Ranjith:2007aa,Padinhateeri:2011aa,Mobius:2013aa,Osberg:2014aa}.
However, in almost all studies, particles have been assumed to interact solely through hard-core steric exclusion, despite the fact that realistic particles typically interact with a soft-core potential due to their internal structure \cite{Baule:2018aa}. In fact, including soft particle interactions significantly alters the filling behaviour and can account, e.g., for the observed rapid filling of nucleosomes on DNA \cite{Mobius:2013aa,Osberg:2014aa,Osberg:2015aa}. In this work, we derive the exact time dependent solution of 1$d$ continuum RSA processes with arbitrary finite-range particle interactions using an iterative approach. This method also provides the exact general solution of the related RSA of polydisperse particles, which has been an open problem since the 1960s {for the continuum 1$d$ case} \cite{Ney:1962aa,Mullooly:1968aa,Krapivsky:1992aa,Brilliantov:1996aa,Brilliantov:1997aa,Burridge:2004aa}, notwithstanding specific scaling solutions that have been found \cite{Krapivsky:1992aa,Brilliantov:1996aa,Brilliantov:1997aa,Hassan:1997aa}. The exact solution allows in particular to address the fundamental question how the packing density of the deposition process can be optimized by tuning the particle interactions or size distribution. Remarkably, the solution reveals that a unique interaction potential/size distribution exists that leads to a maximally dense coverage of the line, which is approached {as $\sim t^{-\nu}$, where $\nu\to 0^+$ is infinitesimally small.}

In RSA, a particle's position is selected with uniform probability over the domain and it is then placed sequentially if there is no overlap with any previously placed particles. Particles are not able to move or reorient once being placed. {In order to parametrize the RSA dynamics of equal particles by a rate equation in $1d$, we define an interval $x$ as the distance between the centers of two nearest-neighbor particles on the line of length $L$ and introduce $N(x,t)\D x$ as the number of intervals with size $\in[x,x+\D x]$ at time $t$. The definition of $x$ implies that $\int_0^L\D x \,x\,N(x,t)=L$ is a conserved quantity of the dynamics for all $t$. The interval distribution $p(x,t)$ is defined as $p(x,t)=\frac{\lambda}{L}N(x,t)$, where $\lambda$ is a characteristic length scale associated with the particles \footnote{For hard particles $\lambda$ would be identified as the size of the particles. For interacting (soft) particles, $\lambda$ could be identified as the interaction range, effective size, or hard core of the particles.}. Scaling length as $x\to x/\lambda$ and considering the limit $L\to\infty$, the time evolution of $p(x,t)$ is exactly described by the master equation}
\be
\label{master}
\frac{\partial}{\partial t}p(x,t)&=&-\psi(x)p(x,t)+2\int_{x}^\infty\D y\,\Omega(x,y)p(y,t),
\ee
where the first term on the rhs describes the destruction of intervals of length $x$ and the second term the creation. $\Omega(x,y)$ is the probability per unit time that a particle is placed inside an interval of length $y$, thus creating an interval of length $x$. The factor 2 stems from the fact that in 1$d$ there are always two ways of doing this for a given $y$ ($x$ is either the distance of the newly inserted particle with the existing left or right particle). The destruction term is accordingly governed by the function $\psi(x)=\int_0^{x}\D u\, \Omega(u,x)$. 
{The interval distribution satisfies $\int_0^\infty\D x \,x\,p(x,t)=1$ for all $t$. Moreover, since one particle is associated with every interval $x$, the integral $\int_0^\infty\D x\,p(x,t)=n(t)$ equals the number density of particles. The requirement of an initially empty line thus leads to the initial condition $n(0)=0$ or $p(x,0)=0$. From Eq.~\eqref{master} this implies that $\lim_{t\to 0}\frac{\partial}{\partial t}p(x,t)=0$, but one can show that $\lim_{t\to 0}\frac{\partial^2}{\partial t^2}p(x,t)>0$ such that $n(t)$ monotonically increases with time. The key quantity of interest is the line coverage (packing density)}
\be
\label{phi}
\phi(t)=1-\int_{\sigma}^\infty\D x\, (x-\sigma)\,p(x,t),
\ee
where $\sigma$ denotes the effective size (diameter) of a particle (see below). {Note that for soft particles $n(t)$ and $\phi(t)$ have very different long-time behaviours: while the jamming density $\phi_{\rm J}=\lim_{t\to\infty}\phi(t)\le 1$, $n(t)$ can diverge as $t\to\infty$, since particles can be absorbed inside existing particles. By contrast, for hard particles they are directly related as $\phi(t)=\sigma n(t)$ \cite{Talbot:2000aa}.} 

The simplest example of a deposition process governed by Eq.~\eqref{master} is Renyi's seminal car parking problem, where particles only interact by steric repulsion \cite{Renyi:1958aa,Renyi:1963aa,Mackenzie:1962aa,Widom:1966aa}. In this case, an exact solution for $p(x,t)$ is known and yields as hallmark features $\phi_{\rm J}= 0.7475...\equiv\phi_{\rm R}$ with the algebraic asymptotic approach $\phi_{\rm J}-\phi(t)\sim t^{-1}$. Other models of the form Eq.~\eqref{master} have been solved for specific $\Omega$ that lead to scale invariant solutions $p$, e.g, RSA \cite{Krapivsky:1992aa,Brilliantov:1996aa,Brilliantov:1997aa} and fragmentation processes \cite{Ziff:1985aa,Cheng:1988aa,Cheng:1990aa,Williams:1990aa}. In the following, an exact analytical solution is derived much more generally without the requirement of scale invariance.

{For interacting particles we assume the form
\be
\label{omega}
\Omega(x,y)=k_+\,\omega(x)\omega(x-y),
\ee
where the adsorption rate $k_+$ sets the time scale (set to unity) and $\omega(x)$ describes the modification in the rate due to particle overlap. The particle interactions are constrained as follows: (i) $\omega(x)=1$ for $x\ge a$, i.e., $a$ is the finite range of the interaction; and (ii) $\omega(x)=0$ for $x\le\Delta$, i.e., $\Delta$ is the hard-core exclusion volume of a particle. Clearly, (i,ii) are satisfied for almost all realistic particle models at least to a good approximation. The hard particle case of Renyi is obtained as $\omega(x)=\Theta(x-a)$ with $\Delta=a$, where $\Theta(x)$ denotes the Heaviside step function. For soft particles $\omega(x)$ interpolates between 0 and 1 (see Fig.~\ref{Fig:pot}a). 
}

The properties (i,ii) of $\omega(x)$ constrain the form of $\psi$ and $\Omega(x,y)$ as follows (Appendix Sec.~\ref{Sec:constraints}):
\be
\label{omegacon}\Omega(x,y)&=&\omega(y-x),\qquad x>a\\
\label{psicon1}\psi(x)&=&0,\qquad\qquad x\le 2\Delta\\
\label{psicon2}\psi(x)&=&x-2\sigma,\qquad x\ge 2a\label{psilinear}
\ee
{where $\sigma=a-\int_\Delta^a\D u\,\omega(u)$ can be interpreted as the effective size of a particle.}

With Eqs.~(\ref{omegacon}--\ref{psicon2}) the solution of Eq.~\eqref{master} can be found as follows. For $x\ge 2a$, Eq.~\eqref{master} simplifies to
\be
\label{master0}
\frac{\partial}{\partial t}p&=&-(x-2\sigma)p+2\int_{x+\Delta}^\infty\D y\,\omega(y-x)p(y,t).
\ee
Crucially, Eq.~\eqref{master0} admits an exact closed form solution. We make the ansatz $p(x,t)=A(t)e^{-(x-2\sigma)t}$, which, upon substitution into Eq.~\eqref{master0}, yields an equation for $A$: $\dot{A}/A=2\int_\Delta^\infty\D u\,\omega(u)e^{-ut}$. In order to make the divergence in the integral at $t=0$ explicit, we perform a partial integration: $\int_\Delta^\infty\D u\,\omega(u)e^{-ut}=t^{-1}\int_\Delta^a\D u\rho(u)e^{-ut}$, where the function $\rho$ is defined as $\rho(x)=\omega'(x)$, which has support on $[\Delta,a]$ and satisfies $\int_\Delta^a\D u\,\rho(u)=1$. The boundary conditions further suggest the form $A(t)=t^2F(t)$ with $F(0)=1$ and we obtain the regularized form
\be
\label{Feq}
F(t)=\exp\left[-2\int_0^t\D s\frac{1-\int_\Delta^a\D u\,\rho(u)e^{-us}}{s}\right].
\ee
Let us denote the resulting solution of Eq.~\eqref{master0} as $p_0$, valid for $x\ge 2a$:
\be
\label{p0sol}
p_0(x,t)=t^2F(t)e^{-(x-2\sigma)t}.
\ee
Eq.~\eqref{p0sol} is valid for arbitrary $\Delta \ge 0$. We now distinguish the two cases $\Delta >0$ and $\Delta =0$.

The key observation is that for a finite excluded volume $\Delta >0$, we directly obtain a solution in the regime $2a-\Delta\le x\le 2a$ (denoted by $p_1$) even though $\psi$ is non-linear: {property (ii) of $\omega(x)$ enforces} the constraint $\Theta(y-x-\Delta)$ in the integral in Eq.~\eqref{master}, which thus only integrates over $p_0${, i.e., the master equation for $2a-\Delta\le x\le 2a$ is
\be
\frac{\partial}{\partial t}p_1(x,t)&=&-\psi(x)p_1(x,t)+2\int_{x+\Delta}^\infty\D y\,\Omega(x,y)p_0(y,t).\nonumber\\
\ee
This simple first-order ODE with inhomogeneity can be directly integrated:
\be
\label{iteration}
p_1(x,t)&=&2\int_0^t\D s\,e^{-\psi(x)(t-s)}\int_{x+\Delta}^\infty\D y\,\Omega(x,y)p_0(y,t).\nonumber\\
\ee
}
Likewise, the solution in the range $2a-2\Delta\le x\le 2a-\Delta$ can be obtained by integration over $p_0$ and $p_1$, and so on. Overall, we thus decompose the interval distribution as
\be
\label{pdecomp}
p(x,t)=\left\{\begin{matrix}p_0(x,t), & x\ge 2a \\ & \\ p_1(x,t), & 2a-\Delta\le x< 2a \\  \vdots & \\ p_j(x,t) & 2a-j\Delta \le x< 2a-(j-1)\Delta \\ \vdots & \\ p_n(x,t) & 2\Delta \le x < 2a-(n-1)\Delta \\ & \\
p_{n+1}(x,t) & \Delta < x <2\Delta \end{matrix}\right.
\ee
Due to the excluded volume, $p(x,t)=0$ for $x\le\Delta$. We introduce the shorthand notation $\Theta_j(x)$ to separate the different $x$ ranges in Eq.~\eqref{pdecomp} and write $p(x,t)=\sum_{j=0}^{n+1}\Theta_j(x)p_j(x,t)$. The solutions $p_1,...,p_{n+1}$ are thus obtained by solving the ODE Eq.~\eqref{master} for each range leading to
\be
\label{iteration}
p_j(x,t)&=&2\int_0^t\D s\,e^{-\psi(x)(t-s)}\int_{x+\Delta}^\infty\D y\,\Omega(x,y)\times\nonumber\\
&&\times\sum_{i=0}^{j-1}\Theta_i(y)p_i(y,s),\qquad j=1,...,n
\ee
and
\be
\label{pn1}
p_{n+1}(x,t)&=&2\int_0^t\D s\int_{x+\Delta}^\infty\D y\,\Omega(x,y)\sum_{i=0}^{n}\Theta_i(y)p_i(y,s),\nonumber\\
\ee
since $\psi(x)=0$ for $x\le 2\Delta$. Eqs.~(\ref{p0sol}--\ref{pn1}) represent the exact analytical solution of Eq.~(\ref{master}) for finite range interactions and $\Delta>0$.

From Eq.~\eqref{Feq} we obtain the asymptotic behaviour $p_0(x,t)\sim A_0 e^{-(x-2\sigma)t}$ and in general $p_j(x,t)\sim A_j(x)e^{-\psi(x) t}$ for $j=1,...,n$, where the $A_j$ can be calculated iteratively (Appendix Sec.~\ref{Sec:time}). We see that the distributions $p_1,...,p_n$ all decay exponentially governed by $\psi(x)$. The stationary limit is thus entirely determined by $p_{n+1}$: $\lim_{t\to\infty}p(x,t)=\lim_{t\to\infty}p_{n+1}(x,t)=p_{\rm s}(x)$, where from Eq.~\eqref{iteration}
\be
\label{pstat}
p_{\rm s}(x)=2\int_0^\infty\D s\,\int_{x+\Delta}^\infty\D y\,\Omega(x,y)\sum_{i=0}^n\Theta_i(y)p_i(y,s).
\ee
With Eq.~\eqref{pstat} the jamming density can be directly evaluated: $\phi_{\rm J}=1-\int_{\sigma}^{2\Delta}\D x (x-\sigma)p_{\rm s}(x)$.

When $\Delta\to 0$, we can use a limiting procedure in Eqs.~(\ref{iteration},\ref{pn1}) or use Eq.~\eqref{p0sol} in Eq.~\eqref{master} and separate the integration region. In both cases, the result is
\be
\label{master2}
\frac{\partial}{\partial t}p_<(x,t)&=&-\psi(x)p_<(x,t)+2\int_x^{2a}\D y\,\Omega(x,y)p_<(y,t)\nonumber\\
&&+2\int_{2a}^\infty\D y\,\Omega(x,y)p_0(y,t),
\ee
where $p_<$ is the interval distribution for the whole range $0\le x\le 2a$. Using the Duhamel principle, Eq.~\eqref{master2} can be solved by iteration leading to a series solution. We define the compact linear operator $\mathcal{L}f(x,t)=2\int_0^t\D s\,e^{-\psi(x)(t-s)}\int_x^{2a}\D y\,\Omega(x,y)f(y,t)$ and obtain the formal solution
\be
p_<(x,t)&=&\frac{1}{1-\mathcal{L}}h(x,t)\nonumber\\
&=&h(x,t)+\mathcal{L}h(x,t)+\mathcal{L}\mathcal{L}h(x,t)+...\label{pdelta0}\\
h(x,t)&=&2\int_0^t\D s\,e^{-\psi(x)(t-s)}\int_{2a}^\infty\D y\,\Omega(x,y)p_0(y,s)\nonumber\label{hdef}\\
\ee
Convergence of this series needs to be established for a given $x,t$ range and $\Omega$.

\begin{figure}
\begin{center}
\includegraphics[width=0.95\columnwidth]{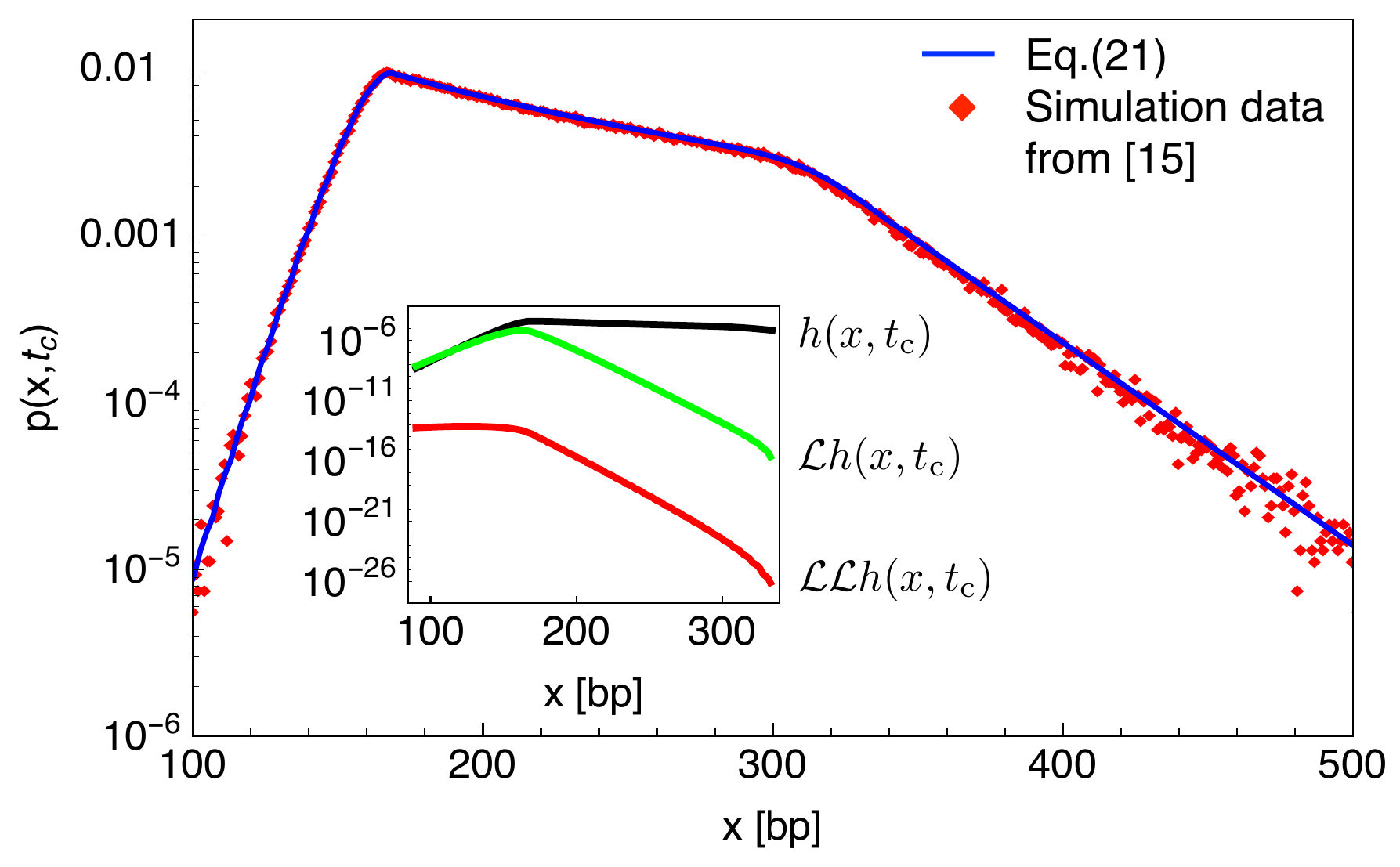}
\caption{\label{Fig:dna}(Colors online) Comparison of the analytical result Eq.~\eqref{dnasol} and the numerically obtained interval distribution at cramming onset $t_{\rm c}$ from simulations of the equilibrium dynamics for the nonlinear potential Eq.~\eqref{dnapot} \cite{Osberg:2014aa}. Since the condition {$\phi(t_{\rm c})=\phi_{\rm R}$} fixes the time scale $k_+$, there is no free parameter in the theory. Inset: The first three terms in the series solution Eq.~\eqref{pdelta0}, indicating convergence after the first two terms.}
\end{center}
\end{figure}

As an example, where these results can provide novel analytical insight, we consider a model of nucleosome packing on DNA. In \cite{Mobius:2013aa,Osberg:2014aa} genome packaging in eukaryotic cells has been modelled assuming an effective `softness' of the nucleosomes due to a multitude of internal states with different footprints on DNA \cite{Mobius:2013aa,Osberg:2014aa}. The effective interaction potential has been approximated from in vivo data as \cite{Mobius:2013aa}
\be
\label{dnapot}
V(x)\approx (a-x)\kappa-\log\left[1+(a-x)\left(1-e^{-\kappa}\right)\right],
\ee
where $a=167$ base pairs (bp) is the maximal footprint size and $\kappa=0.15$ the stiffness per bp. A numerical investigation of the equilibrium kinetics of Eq.~\eqref{dnapot} {exhibits a universal interval distribution independent of the adsorption rate $k_+$ at a specific time $t_{\rm c}$. The time $t_{\rm c}$ is defined by the condition $\phi(t_{\rm c})=\phi_{\rm R}$ and denotes the onset of `cramming', where nucleosomes are increasingly squeezed into gaps that are smaller than the maximal footprint size \cite{Mobius:2013aa,Osberg:2014aa}. The dynamics in the cramming regime can be described as follows. In nucleosome packing both adsorption (with rate $k_+$) and desorption (with rate $k_-$) occur whereby $k_+\gg k_-$. Thus in a time regime $\ll 1/k_-$ the equilibrium dynamics is well described by an irreversible RSA process like Eq.~\eqref{master} \cite{Epstein:1979aa,Epstein:1979ab}. Here, the effect of the interaction potential can be captured by the Boltzmann factor \cite{Mobius:2013aa,Osberg:2014aa}
\be
\label{omegaV}
\Omega(x,y)=k_+\exp\left[-V(x)-V(y-x)\right],
\ee
setting $k_BT=1$ and the normalization constant is included in $k_+$. Since $t_{\rm c}$ has been observed to be $\ll 1/k_-$, the RSA process applies in the cramming regime and explains immediately the apparent universality of the interval distribution: different $k_+$ lead to different $t_{\rm c}$ values, but the resulting $p(x,t_{\rm c})$ are all identical, since they are determined at the same $\phi$ value. In fact, this observation shows that the universality holds not only at cramming onset, but for any fixed $\phi$ value in the regime $\ll 1/k_-$. It also highlights that the curves $\phi(t)$ for different $k_+$ are all scaled versions of each other up to times $t\approx 1/k_-$, which is indeed suggested in the numerical results of \cite{Osberg:2014aa}. Remarkably, with Eqs.~(\ref{p0sol},\ref{pdelta0},\ref{hdef}) we obtain
\be
\label{dnasol}
p(x,t_{\rm c})=\left\{\begin{matrix} p_0(x,t_{\rm c}),&x\ge 2a \\ (1+\mathcal{L})h(x,t_{\rm c}),& 0\le x<2a \end{matrix}\right.
\ee
which shows perfect agreement with the numerically obtained interval distribution at $t_{\rm c}$ \cite{Osberg:2014aa} despite the strong nonlinear character of the interaction potential Eq.~\eqref{dnapot}.}

We now want to understand how dense packings on the line can be generated by tuning the interaction potential $V$ {assuming $\omega(x)=e^{-V(x)}$ as in Eq.~\eqref{omegaV}.} For simplicity, we restrict the discussion to purely repulsive interactions such that $\omega(x)$ is monotonically increasing with $0\le \omega(x)\le 1$ for $x\in [\Delta,a]$. {We then make two key observations: (a) For $\sigma\ge 2\Delta$ the effective size is larger than the minimal separation of two particles, thus the line will eventually be fully covered by particles and $\phi_{\rm J}=1$. We thus assume the more interesting case $2\Delta>\sigma$ in the following. (b) For $a>\Delta$ a potential leading to a maximum in $\phi_{\rm J}$ must exist. We can conclude this surprising fact from the two limiting forms of $\omega$ that satisfy the properties (i,ii) and the repulsive nature (see Fig.~\ref{Fig:pot}a). One limit is $\omega(x)=\Theta(x-a)$ (leading to $\sigma=a$) and the other limit is $\omega(x)=\Theta(x-\Delta)$ (leading to $\sigma=\Delta$). However, in both cases we recover $\phi_{\rm J}=\phi_{\rm R}$, since $\phi_{\rm J}$ is invariant with respect to a single size scale $a$ or $\Delta$ on an infinite line.} When $a>\Delta$, we have $\phi_{\rm J}>\phi_{\rm R}$, so a potential leading to a maximum must exist for a given $a$. What is the form of this optimal potential?

In order to elucidate this matter, we consider a one-parameter family of potentials that can interpolate between the two limiting forms of $\omega(x)$ (see Fig.~\ref{Fig:pot}a)
\be
\label{mupot}
V(x)=\left\{\begin{matrix}-\mu\log\left(\frac{x-\Delta}{a-\Delta}\right),&\Delta\le x\le a \\ 0, & x>a \end{matrix}\right.
\ee
where $\mu>0$. {The resulting $\phi_{\rm J}$ increases monotonically for larger $\mu$} (Fig.~\ref{Fig:pot}c). Surprisingly, this suggests that the maximally dense packing is reached when $\omega(x)$ becomes infinitesimally close to the step function limit $\Theta(x-a)$. One can show (Appendix Sec.~\ref{Sec:asymp}) that the interval distribution then converges to the stationary limit $p_{\rm s}(x)=\Theta(x-a)\Theta(2\Delta-x)\tilde{p}^*(x-a)$, where
\be
\label{ptildestat}
\tilde{p}^*(z)=2\int_0^\infty\D t\,t\,\exp\left[-zt-2\int_0^t\D s\frac{1-e^{-as}}{s}\right],
\ee
and the corresponding maximal line coverage is
\be
\label{maxphi}
\phi_{\rm opt}=\phi_{\rm R}+\int_{2\Delta-a}^a\D u\,z\,\tilde{p}^*(z).
\ee
As shown in Fig.~\ref{Fig:pot}b, $p_{\rm s}(x)$ diverges at $x=\sigma=a$ indicating a large number of particle configurations with no empty space between neighbors. The optimal density $\phi^{\rm opt}$ increases monotonically as a function of $a$ (see inset of Fig.~\ref{Fig:pot}c). Note that the limit $\mu\to\infty$ is singular, recovering the Renyi density $\phi_{\rm R}$ instead (Appendix Sec.~\ref{Sec:asymp}). Remarkably, Eqs.~(\ref{ptildestat},\ref{maxphi}) remain valid for any potential for which $\omega(x)$ approaches $\Theta(x-a)$ infinitesimally closely, highlighting that these results are independent of the specific form Eq.~\eqref{mupot}. This non-trivial property can be understood intuitively. In the $\mu\gg 1$ regime the system behaves initially like the Renyi car parking of hard particles: since $\omega(x)$ is infinitesimally close to $\Theta(x-a)$ all trial configurations with particle overlap are practically rejected until no gaps $x\ge 2a$ remain. {Subsequently, the steep decay of $\omega(x)$ prevents that intervals $x<2a$ are filled with larger than necessary overlaps. 
The optimal potential thus induces a perfectly hierarchical filling, where all intervals $2\Delta<x<2a$ are eventually filled, but extremely slowly.} 
In fact, with the previous results and the asymptotic properties of $\psi$ we find that for $\mu\gg 1$ (Appendix Sec.~\ref{Sec:asymp2})
\be
\phi_{\rm J}-\phi(t)\sim\int_{2\Delta}^{2a-(n-1)\Delta}\D x\, e^{-\psi(x) t}\sim t^{-\nu},
\ee
where $\nu=\frac{1}{2\mu+1}$. Nevertheless, Eq.~(\ref{maxphi}) can be verified in a simulation by running the Renyi car parking RSA with length scale $a$ and then considering all intervals $x>2\Delta$ as filled when evaluating $\phi_{\rm J}$. This indeed yields perfect agreement with the theory (see inset of Fig.~\ref{Fig:pot}c).

\begin{figure}
\begin{center}
\includegraphics[width=1.0\columnwidth]{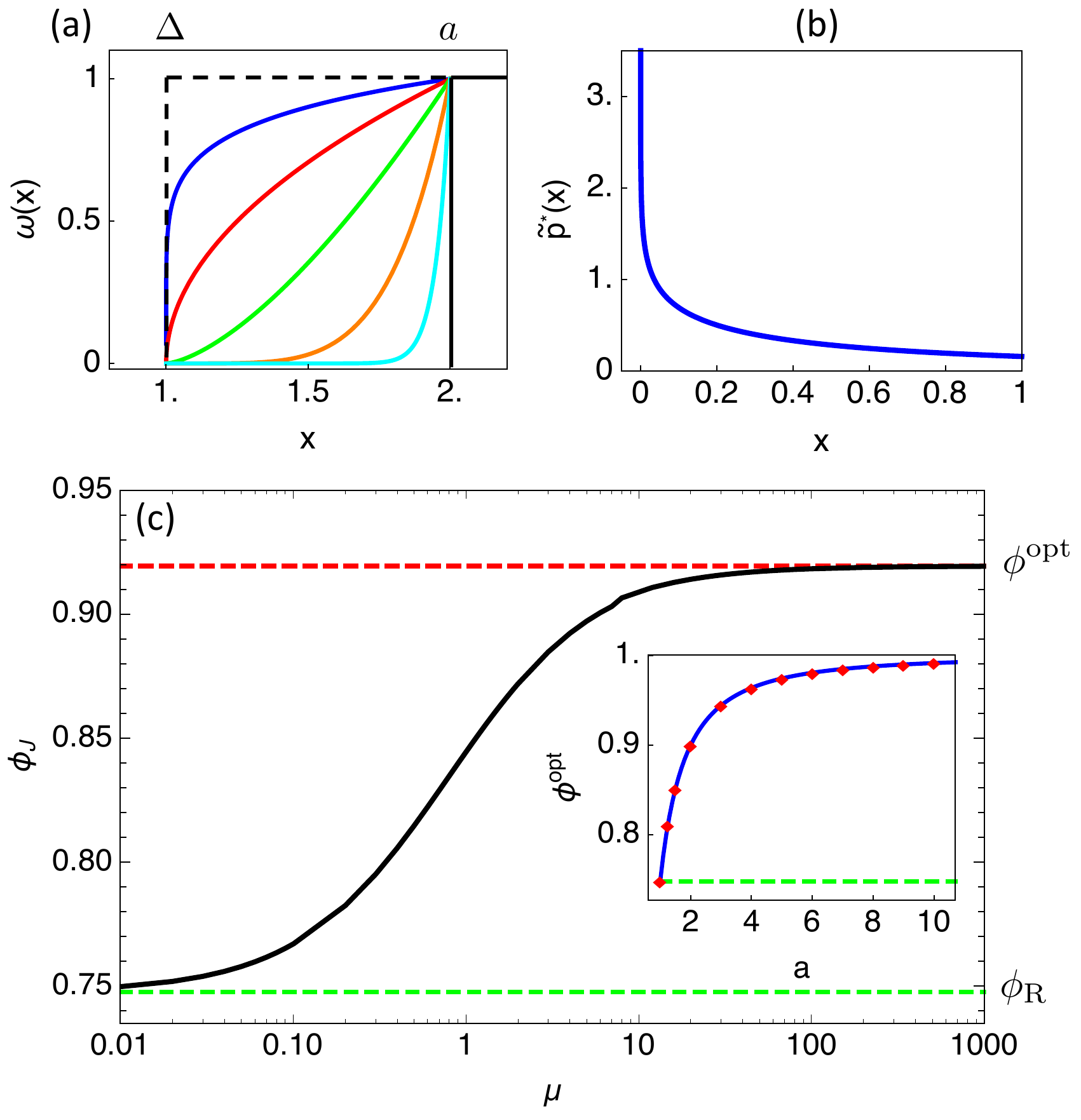}
\caption{\label{Fig:pot}(Colors online) (a) The two limiting forms of $\omega(x)$: $\Theta(x-\Delta)$ and $\Theta(x-a)$ together with interpolations from the potential Eq.~\eqref{mupot}. (b) Plot of $\tilde{p}^*(z)$, Eq.~\eqref{ptildestat}, the distribution of empty gaps ($z=x-a$) for $t\to\infty$ and $\mu\gg 1$ (here $a=1$). (c) Plot of $\phi_{\rm J}$ resulting from the potential Eq.~\eqref{mupot} calculated from the analytical solution Eqs.~(\ref{p0sol}--\ref{pn1}) (here $a=1.4$, $\Delta=1$). Inset: Plot of $\phi^{\rm opt}$, Eq.~\eqref{maxphi}, together with simulation results (here $2\Delta-a=1$).}
\end{center}
\end{figure}

The iterative solution method can be applied to related interacting particle models such as the {continuum} RSA of polydisperse particles in 1$d$, whose general time-dependent solution has been a long-standing open problem \cite{Ney:1962aa,Mullooly:1968aa,Krapivsky:1992aa,Brilliantov:1996aa,Brilliantov:1997aa,Burridge:2004aa}. {Crucially, this problem can be solved in full generality akin to the interacting particle case for an arbitrary size distribution $\rho(\sigma)$ with support $\sigma\in[\sigma_1,\sigma_2]$ (Appendix Sec.~\ref{Sec:poly}).}
The solution shows in particular that the exponent of the asymptotic approach is determined by the leading term in the asymptotic expansion of $\rho(\sigma)$ as $\sigma\to\sigma_1$, which makes sense intuitively because the large time asymptotics is determined by the filling of the smallest available intervals.
The widely used assumption that polydispersity generally reduces the exponent by an additional degree of freedom from $\nu=1/d$ to $\nu=1/(d+1)$ \cite{Tarjus:1991ac,Evans:1993ab,Adamczyk:1997aa,Talbot:2000aa}, where $d$ is the spatial dimension, is thus not correct for $d=1$. A similar result should hold for the RSA of polydisperse spheres in higher dimensions, where the quantitative dependence on $\rho$ has first been observed empirically \cite{Meakin:1992aa}. {Remarkably, the size distribution $\rho(x)$ that leads to a maximal line coverage yields the same interval distribution and optimal density as in the interacting particle case, given by Eqs.~(\ref{ptildestat},\ref{maxphi}), where now $\sigma_1=2\Delta-a$ and $\sigma_2=a$ (Appendix Sec.~V).}

{The RSA of interacting particles coresponds to a cooperative RSA problem, where adsorption rates depend on the local environment of the particle with a given range \cite{Evans:1993ab}. Cooperative RSA problems have long been studied on discrete lattices, for which exact results are available for $M$-mers (particles occupying $M$ sites) with cooperativity of range-1 \cite{Gonzalez:1974aa,Boucher:1973aa,Epstein:1979aa}, range-$M$ \cite{Wolf:1984aa,Mellein:1986ab}, and arbitrary finite range-$R$ \cite{Evans:1990aa}. From such discrete models, the continuum equivalent is obtained in the limit of $M\to\infty$. However, performing such a limit in the general case of $M$-mers with general range-$R$ cooperativity, which could be mapped onto the continuum RSA of interacting particles, is non-trivial and left for future work.} Further applications of the exact results in the context of nucleosome packing models would be highly interesting. The empirically obtained potential Eq.~\eqref{dnapot}, e.g., might represent an optimal trade-off between achieving a high coverage and a fast filling dynamics on time scales relevant for biological function. To this end not only $\phi_{\rm J}$ but the dynamics of $\phi(t)$ needs to be explored in the space of possible potentials. In \cite{Gutierrez:2015aa}, cold atoms that are excited to high-lying Rydberg states are shown to fill up excitation levels akin to a classical deposition process, where atoms interact through a highly nonlinear potential with excluded volume \cite{Lesanovsky:2013aa}. The analytical solution obtained here could provide important deeper insight into such cold atom excitations, which would be readily realizable in experiments.

\begin{acknowledgments}

AB gratefully acknowledges funding under EPSRC grant EP/L020955/1 and helpful discussions with R. Guti\'errez, J.~P. Garrahan, and I. Lesanovsky. AB thanks B. Osberg, J. N\"ubler, and U. Gerland for sharing the simulation data of Fig.~4A in \cite{Osberg:2014aa}.

\end{acknowledgments}

\onecolumngrid

\begin{appendix}

\section{Constraints on the master equation for finite-range interactions}
\label{Sec:constraints}

Including explicitly the constraints (i,ii) on $\omega(x)$, we can express $\psi(x)=\int_0^x\D y\,\omega(x-y)\omega(y)$ as
\be
\psi(x)&=&\int_0^x\D y\,[\Theta(y-\Delta)\Theta(a-y)\omega(y)+\Theta(y-a)] [\Theta(x-y-\Delta)\Theta(a-x+y)\omega(x-y)+\Theta(x-y-a)].
\ee
Factorization yields four terms
\be
\label{psifact}
\psi(x)&=&\psi_1(x)+\psi_2(x)+\psi_3(x)+\psi_4(x)\\
\psi_1(x)&=&\int_0^{x}\D y\Theta(a-y)\Theta(y-\Delta)\Theta(a-x+y)\Theta(x-y-\Delta)\omega(y)\omega(x-y)\nonumber\\
&=&\Theta(x-2\Delta)\Theta(a+\Delta-x)\int_\Delta^{x-\Delta}\D y\,\omega(y)\omega(x-y)\nonumber\\
&&+\Theta(2a-x)\Theta(x-(a+\Delta))\int^a_{x-a}\D y\,\omega(y)\omega(x-y)\nonumber\\
\psi_2(x)&=&\int_0^x\D y\,\Theta(y-\Delta)\Theta(a-y)\Theta(x-y-a)\omega(y)\nonumber\\
&=&\Theta(x-(a+\Delta))\Theta(2a-x)\int_\Delta^{x-a}\D y\,\omega(y)+\Theta(x-2a)\int_\Delta^{a}\D y\,\omega(y)\nonumber\\
\psi_3(x)&=&\int_0^x\D y\,\Theta(y-a)\Theta(x-y-\Delta)\Theta(a-x+y)\omega(x-y)\nonumber\\
&=&\Theta(x-(a+\Delta))\Theta(2a-x)\int_a^{x-\Delta}\D y\,\omega(x-y)+\Theta(x-2a)\int^{x-\Delta}_{x-a}\D y\,\omega(x-y)\nonumber\\
\psi_4(x)&=&\int_0^x\D y\,\Theta(y-a)\Theta(x-y-a)\nonumber\\
&=&\Theta(x-2a)(x-2a)\nonumber
\ee
We see immediately that $\psi(x)=0$ for $x\le 2\Delta$, i.e., Eq.~\eqref{psicon1} holds. On the other hand, for $x\ge 2a$, the first term in Eq.~\eqref{psifact} vanishes and the second and third terms both become $\int_\Delta^{a}\D y\,\omega(y)$, which leads to the linear form Eq.~\eqref{psicon2}.

\section{Asymptotic time regime}
\label{Sec:time}

We determine successively expressions for $p_0,p_1,...$ that are valid as $t\to\infty$. First, using partial integration in Eq.~\eqref{Feq} we obtain
\be
\int_0^t\D s\left(\frac{1-\langle e^{-us}\rangle}{s}\right)&=&\left(1-\langle e^{-us}\rangle\right)\log(t)-\int_0^t\D s\langle u e^{-us}\rangle\log(s)
\ee
where we use the shorthand notation $\langle...\rangle=\int_\Delta^a\D u\,...\rho(u)$. Asymptotically, $p_0$ of Eq.~\eqref{p0sol} becomes for large $t$
\be
\label{p0asymp}
p_0(x,t)\simeq A_0e^{-(x-2\sigma)t},\qquad A_0=e^{2\int_0^\infty\D s\langle u e^{-us}\rangle\log(s)}.
\ee
In order to evaluate $p_1$, we note that Eq.~\eqref{iteration} for $j=1$ can be simplified since the integrations only act on $p_0$ and we can substitute Eq.~\eqref{master0}
\be
p_1(x,t)&=&\int_0^t\D s\,e^{-\psi(x)(t-s)}\left(\frac{\partial}{\partial s}+(x-2\sigma)\right)p_0(x,t)\nonumber\\
&=&p_0(x,t)+(x-2\sigma-\psi(x))\int_0^t\D s\,e^{-\psi(x)(t-s)}p_0(x,s).\label{p1asymp}
\ee
If we introduce a time $t_{\rm a}$ such that the asymptotic behaviour Eq.~\eqref{p0asymp} is valid for $t>t_{\rm a}$, we obtain from Eq.~\eqref{p1asymp} by separating the integral
\be
p_1(x,t)&=&p_0(x,t)+e^{-\psi(x)(t-t_{\rm a})}(p_1(x,t_{\rm a})-p_0(x,t_{\rm a}))+(x-2\sigma-\psi(x))\int_{t_{\rm a}}^t\D s\,e^{-\psi(x)(t-s)}A_0e^{-(x-2\sigma)s}\nonumber\\
&=&p_0(x,t)+e^{-\psi(x)(t-t_{\rm a})}(p_1(x,t_{\rm a})-p_0(x,t_{\rm a}))+A_0e^{-(x-2\sigma)t_{\rm a}}e^{-\psi(x)(t-t_{\rm a})}-A_0e^{-(x-2\sigma)t}\nonumber\\
&\simeq&p_1(x,t_{\rm a})e^{-\psi(x)(t-t_{\rm a})}.
\ee
We thus obtain the asymptotic behaviour for $t\to\infty$
\be
\label{p1asymp1}
p_1(x,t)&\simeq&A_1(x)e^{-\psi(x)t},\qquad A_1(x)=p_1(x,t_{\rm a})e^{\psi(x)t_{\rm a}}.\nonumber\\
\ee
From Eq.~\eqref{iteration} we obtain likewise for $p_2$
\be
p_2(x,t)&=&e^{-\psi(x)(t-t_{\rm a})}p_2(x,t_{\rm a})+2\int_{t_{\rm a}}^t\D s\,e^{-\psi(x)(t-s)}\left[\int_{x+\Delta}^{2a}\D y\,\Omega(x,y)A_1(y)e^{-\psi(y)s}+\int_{2a}^\infty\D y\,\Omega(x,y)A_0e^{-(y-2\sigma)s}\right]\nonumber\\
&=&e^{-\psi(x)(t-t_{\rm a})}p_2(x,t_{\rm a})+\int_{x+\Delta}^{2a}\D y\,\frac{\Omega(x,y)A_1(y)}{\psi(y)-\psi(x)}\left[e^{-\psi(y)t_{\rm a}-\psi(x)(t-t_{\rm a})}-e^{-\psi(y)t}\right]\nonumber\\
&&+A_0\int_{2a}^\infty\D y\,\frac{\Omega(x,y)}{y-2\sigma-\psi(x)}\left[e^{-(y-2\sigma)t_{\rm a}-\psi(x)(t-t_{\rm a})}-e^{-(y-2\sigma)t}\right]
\ee
The only problematic term is here the integral
\be
\label{unknown}
\int_{x+\Delta}^{2a}\D y\,\frac{\Omega(x,y)A_1(y)}{\psi(y)-\psi(x)}e^{-\psi(y)t},
\ee
whose precise asymptotics for $t\to\infty$ can not be determined in general, since $\Omega(x,y)$ vanishes exponentially fast at $y=x+\Delta$, the minimum of $\psi(y)$. We assume that this term decays faster than $e^{-\psi(x)t}$, since $\psi(x+\Delta)>\psi(x)$ in general, but this needs to be argued for each potential $V$. With this assumption we obtain for $t>t_{\rm a}$
\be
p_2(x,t)&\simeq& A_2(x)e^{-\psi(x)t}\\
A_2(x)&=&p_2(x,t_{\rm a})e^{\psi(x)t_{\rm a}}+\int_{x+\Delta}^{2a}\D y\,\frac{\Omega(x,y)A_1(y)}{\psi(y)-\psi(x)}e^{-(\psi(y)-\psi(x))t_{\rm a}}+A_0\int_b^\infty\D y\,\frac{\Omega(x,y)}{y-2\sigma-\psi(x)}e^{-(y-2\sigma-\psi(x))t_{\rm a}}.\nonumber\\
\ee
The calculation can be continued for general $p_j$, $j=3,...,n$, which yields
\be
\label{pjasymp}
p_j(x,t)&\simeq& A_j(x)e^{-\psi(x)t},
\ee
where $A_j$ depends on $A_0,A_1,...,A_{j-1}$. Even though the derivation of relies on assumptions of the decay of integrals of the type of Eq.~\eqref{unknown}, Eq.~\eqref{pjasymp} is also motivated heuristically from the filling dynamics of Eq.~\eqref{master}: larger intervals are more easily filled than smaller ones, such that the $p_j$ need to decay hierarchically for $t\to\infty$. Since for finite $\Delta$ the creation term in Eq.~\eqref{master} depends on $p_0,p_1,...,p_{j-1}$, it thus vanishes before $p_j$, leading to the asymptotic behaviour Eq.~\eqref{pjasymp}.

\section{Analysis of the $\mu\gg 1$ regime}
\label{Sec:asymp}

Since we are interested in the jamming density $\phi_{\rm J}=\lim_{t\to\infty}\phi(t)$, which is determined from Eq.~\eqref{phi}, we need to clarify the role of the two limits $t\to\infty$ and $\mu\to\infty$. First we note that for the potential Eq.~\eqref{mupot} the following limits apply:
\be
\begin{matrix}
\lim_{\mu\to 0}\omega(x)=\Theta(x-\Delta),\qquad & \lim_{\mu\to 0}\psi(x)=\Theta(x-2\Delta),\qquad & \lim_{\mu\to 0}\sigma=\Delta,\qquad & \lim_{\mu\to 0}\left<e^{-\sigma\,t}\right>=e^{-\Delta t}\\
\lim_{\mu\to \infty}\omega(x)=\Theta(x-a),\qquad & \lim_{\mu\to \infty}\psi(x)=\Theta(x-2a),\qquad & \lim_{\mu\to \infty}\sigma=a,\qquad & \lim_{\mu\to \infty}\left<e^{-\sigma\,t}\right>=e^{-a\, t}\label{mulimits}
\end{matrix}
\ee
In both limits the resulting jamming density is then just the Renyi density $\phi_{\rm R}$, since the particles are effectively hard with only a single length scale. For $\mu\to 0$ this follows formally from Eqs.~(\ref{phi},\ref{pdecomp},\ref{pn1}) as
\be
\phi_{\rm J}&=&1-\lim_{t\to\infty,\mu\to 0}\int_{\sigma}^{\infty}\D x (x-\sigma)p(x,t)\nonumber\\
&=&1-\int_{\Delta}^{2\Delta}\D x (x-\Delta)\lim_{t\to\infty,\mu\to 0}p_{n+1}(x,t)\nonumber\\
&=&1-\int_{\Delta}^{2\Delta}\D x (x-\Delta)2\int_0^\infty\D s\int_{x+\Delta}^\infty\D y\,\lim_{\mu\to 0}\sum_{i=0}^{n}\Theta_i(y)p_i(y,s)\nonumber\\
&=&1-\int_{\Delta}^{2\Delta}\D x (x-\Delta)2\int_0^\infty\D s\int_{x+\Delta}^\infty\D y\,\lim_{\mu\to 0}p_0(y,s)
\ee
where the last step follows since for $\omega(x)=\Theta(x-\Delta)$ all distributions $p_1,...,p_n$ reduce to $p_0$. Using $\lim_{\mu\to 0}p_0(x,t)=t^2\exp\left[-2\int_0^t\D s\frac{1-e^{-\Delta\, s}}{s}\right]e^{-(x-2\Delta)t}$, see Eq.~\eqref{p0sol}, we obtain
\be
\phi_{\rm J}&=&1-\int_{0}^{\Delta}\D x\,2x\int_0^\infty\D t\,t\exp\left[-2\int_0^t\D s\frac{1-e^{-\Delta\, s}}{s}\right]e^{-xt}\label{renyips}\\
&=&1-2\int_0^\infty\D t\,\exp\left[-2\int_0^t\D s\frac{1-e^{-\Delta\, s}}{s}\right]\left(\frac{1-e^{-\Delta t}}{t}-\Delta e^{-\Delta t}\right)\nonumber\\
&=&2\Delta\int_0^\infty\D t\,\exp\left[-2\int_0^t\D s\frac{1-e^{-\Delta\, s}}{s}\right] e^{-\Delta t}\nonumber\\
&=&\Delta\int_0^\infty\D t\,\exp\left[-2\int_0^t\D s\frac{1-e^{-\Delta\, s}}{s}\right]\nonumber\\
&=&\phi_{\rm R}.\label{deltacalc}
\ee
For the limit $\mu\to\infty$, we need to be more careful with the sequence of limits to recover $\phi_{\rm R}$. In particular, if we take $\mu\to\infty$ before $t\to\infty$, the distributions $p_1,...,p_n$ no longer decay to zero for $t\to\infty$, because $\psi(x)=0$ for $x\le 2a$. In fact, Eqs.~\eqref{iteration} then take the form of Eq.~\eqref{pn1}, i.e., $p_1,...,p_n$ recover $p_{n+1}$ whose range of validity is now extended to the interval $[\Delta,2a]$. We thus obtain
\be
\phi_{\rm J}&=&1-\lim_{t\to\infty,\mu\to \infty}\int_{\sigma}^{\infty}\D x (x-\sigma)p(x,t)\nonumber\\
&=&1-\int_{a}^{2a}\D x (x-a)\lim_{t\to\infty,\mu\to \infty}\sum_{i=1}^{n}\Theta_i(x)p_i(x,t)\nonumber\\
&=&1-\int_{a}^{2a}\D x (x-a)\lim_{t\to\infty,\mu\to \infty}p_{n+1}(x,t)\nonumber\\
&=&1-\int_{a}^{2a}\D x (x-a)2\int_0^\infty\D s\int_{x+a}^\infty\D y\,\lim_{\mu\to \infty}p_0(y,s).
\ee
From then on the same calculation as above in Eq.~\eqref{deltacalc} applies, where $\Delta$ is now replaced by $a$ leading also to $\phi_{\rm R}$. What happens now, when $\mu\gg 1$ but without taking the actual limit $\mu\to\infty$? For any large but finite $\mu$, the  distributions $p_1,...,p_n$ will eventually decay to zero for $t\to\infty$, since $\psi(x)$ is vanishingly small but non-zero for $x\le 2a$. Apart from this crucial difference, the other limits Eqs.~\eqref{mulimits} hold effectively. As a result, Eq.~\eqref{phi} becomes
\be
\phi_{\rm J}&=&1-\int_{a}^{2\Delta}\D x (x-a)\lim_{t\to\infty,\mu\to\infty}p_{n+1}(x,t)\nonumber\\
&=&1-\int_{a}^{2\Delta}\D x (x-a)2\int_0^\infty\D s\int_{x+a}^\infty\D y\,\lim_{\mu\to \infty}p_0(y,s)\nonumber\\
&=&1-\int_{0}^{2\Delta-a}\D z\,2z\int_0^\infty\D t\,t\exp\left[-zt-2\int_0^t\D s\frac{1-e^{-as}}{s}\right]\nonumber\\
&=&1-\int_{0}^{2\Delta-a}\D z\,z\,\tilde{p}^*(z).
\ee
Here, $\tilde{p}^*(z)$ is given in Eq.~\eqref{ptildestat}. We see that the stationary distribution is given by
\be
p_{\rm s}(x)&=&\lim_{t\to\infty,\mu\to\infty}p_{n+1}(x,t)=\Theta(x-a)\Theta(2\Delta-x)\tilde{p}^*(x-a).
\ee 
Noting that $\Theta(a-z)\tilde{p}^*(z)$ is the distribution of empty intervals in the Renyi car parking problem (see Eq.~\eqref{renyips} with $\Delta=a$), the jamming density can be further expressed as
\be
\phi_{\rm J}&=&1-\int_0^{2\Delta-a}\D z\,z\,\tilde{p}^*(z)\nonumber\\
&=&1-\int_0^a\D z\,z\,\tilde{p}^*(z)+\int_{2\Delta-a}^a\D z\,z\,\tilde{p}^*(z)\nonumber\\
&=&\phi_{\rm R}+\int_{2\Delta-a}^a\D z\,z\,\tilde{p}^*(z)\nonumber\\
&=&\phi^{\rm opt}.
\label{phijamapp}
\ee
The optimal jamming density $\phi^{\rm opt}$ is thus achieved for arbitrarily large but finite $\mu\gg 1$. As discussed above, the limit $\mu\to\infty$ is singular, recovering instead the Renyi density $\phi_{\rm R}$.

\section{Asymptotic approach to the jamming density}
\label{Sec:asymp2}

The asymptotic approach to $\phi_{\rm J}$ can be calculated with the large $t$ property Eq.~\eqref{pjasymp}. With Eq.~\eqref{phi}, we obtain
\be
\phi_{\rm J}-\phi(t)&=&\phi_{\rm J}-1+\int_{\sigma}^\infty\D x(x-\sigma)\sum_{i=0}^{n+1}\Theta_i(x)p_i(x,t)\nonumber\\
&\simeq&\phi_{\rm J}-1+\int_{\sigma}^{2\Delta}\D x(x-\sigma)p_{n+1}(x,t)+\int_{2\Delta}^{2a-(n-1)\Delta}\D x(x-\sigma)p_n(x,t)\nonumber\\
&\simeq&\int_{2\Delta}^{2a-(n-1)\Delta}\D x(x-\sigma)A_n(x)e^{-\psi(x)t},\label{phijasymp}
\ee
for large $t$, using Eq.~\eqref{pjasymp} and the fact that $p_{n+1}$ converges to the stationary interval distribution, which determines $\phi_{\rm J}$. It remains to determine the behaviour of $\psi(x)$ in the vicinity of the minimum of the integration range at $x=2\Delta$. From Eq.~\eqref{psifact} we obtain as $x\to2\Delta$
\be
\psi(x)&=&\int_\Delta^{x-\Delta}\D y\,\omega(y)\omega(x-y)\sim (x-2\Delta)^{2\mu+1},
\ee
for the potential Eq.~\eqref{mupot}. Since there are no other singularities in the integrand of Eq.~\eqref{phijasymp}, we obtain the asymptotic scaling
\be
\phi_{\rm J}-\phi(t)&\sim&\int_{2\Delta}^{2a-(n-1)\Delta}\D x\,e^{-(x-2\Delta)^{2\mu+1}t}\nonumber\\
&\sim&t^{-\frac{1}{2\mu+1}}.
\ee

\section{Exact solution of the polydisperse RSA problem}
\label{Sec:poly}

We now consider a different RSA problem in 1$d$, where the particles are hard, but their sizes (diameters) $\sigma$ are drawn from an independent distribution $\rho(\sigma)$ at each adsorption attempt. We assume the PDF $\rho$ normalized with finite support $\sigma\in[\sigma_1,\sigma_2]$. In this case, the time evolution of the interval distribution $p(x,t)$ is not described by Eq.~\eqref{master}, since the particles bounding every interval are of different size, which is not taken into account in Eq.~\eqref{master}. Formally, one would need to extend $p(x,t)$ to $p(x,t;\sigma_{\rm L},\sigma_{\rm R})$, where $\sigma_{\rm L},\sigma_{\rm R}$ denotes the sizes of the left and right particles separated by $x$. The time evolution of $p(x,t;\sigma_{\rm L},\sigma_{\rm R})$ follows a more complicated master equation, which can be mapped to the one of the `Paris car parking problem' [35]. Nevertheless, we can describe the polydisperse RSA problem by an equation of the type of Eq.~\eqref{master}, if we focus on the distribution of empty gaps, $\tilde{p}(z,t)$, instead of the interval distribution $p(x,t)$ ($x$ is the distance between the centers of two particles, while $z$ is the uncovered part of $x$). The master equation~\eqref{master} is then
\be
\label{polyms}
\frac{\partial}{\partial t}\tilde{p}(z,t)&=&-\left<(z-\sigma)\Theta(z-\sigma)\right>\tilde{p}(z,t)+2\left<\int_{z+\sigma}^\infty\D y\,\tilde{p}(y,t)\right>,
\ee
where the brackets $\langle...\rangle$ denote an expected value with respect to the size distribution $\rho$. The destruction term takes into account the possible ways of inserting a particle with size $\sigma$ sampled from $\rho$ in an empty gap of length $z$. Likewise the creation term is averaged over all possible sizes. As before, the rate constant is absorbed in the time scale and set to unity. We rewrite
\be
\left<\int_{z+\sigma}^\infty\D y\,\tilde{p}(y,t)\right>&=&\int_{\sigma_1}^{\sigma_2}\D\sigma\,\rho(\sigma)\int_{z+\sigma}^\infty\D y\,\Theta(y-z-\sigma)\tilde{p}(y,t)\nonumber\\
&=&\int_{z+\sigma_1}^\infty\D y\int_{\sigma_1}^{y-z}\D\sigma\,\rho(\sigma)\tilde{p}(y,t)
\ee
and, by comparing with Eqs.~(\ref{master},\ref{omegacon}), identify
\be
\label{psipolydef}
\psi(z)&=&\left<(z-\sigma)\Theta(z-\sigma)\right>\\
\omega(z)&=&\int_{\sigma_1}^{z}\D\sigma\,\rho(\sigma)=\left<\Theta(z-\sigma)\right>,\label{omegapoly}
\ee
where due to normalization $\omega(z)=1$ for $z\ge \sigma_2$. In analogy to Eqs.~(\ref{psicon1},\ref{psicon2}), $\tilde{\psi}(z)$ satisfies
\be
\psi(z)&=&0,\qquad\qquad z\le\sigma_1\\
\psi(z)&=&z-\overline{\sigma},\qquad\qquad z\ge\sigma_2
\ee
where $\overline{\sigma}=\left<\sigma\right>$. As a consequence, the same solution method as in the interacting particle case can be applied. The results equivalent to Eqs.~(\ref{p0sol}--\ref{pn1}) are
\be
\tilde{p}(z,t)=\left\{\begin{matrix}\tilde{p}_0(z,t), & z\ge \sigma_2 \\ & \\ \tilde{p}_1(z,t), & \sigma_2-\sigma_1\le z< \sigma_2 \\  \vdots & \\ \tilde{p}_j(z,t) & \sigma_2-j\sigma_1 \le z< \sigma_2-(j-1)\sigma_1 \\ \vdots & \\ \tilde{p}_n(z,t) & \sigma_1 \le z < \sigma_2-(n-1)\sigma_1 \\ & \\
\tilde{p}_{n+1}(z,t) & 0 \le z <\sigma_1 \end{matrix}\right.
\ee
where
\be
\tilde{p}_0(z,t)&=&t^2F(t)e^{-(z-\overline{\sigma})t},\qquad F(t)=\exp\left[-2\int_0^t\D s\frac{1-\langle e^{-\overline{\sigma} s}\rangle}{s}\right]\label{ptilde0}\\
\tilde{p}_j(z,t)&=&2\int_0^t\D s\,e^{-\psi(z)(t-s)}\int_{z+\sigma_1}^\infty\D y\,\omega(y-z)\sum_{i=0}^{j-1}\Theta_i(y)\tilde{p}_i(y,s),\qquad j=1,...,n\label{pjpoly}\\
\tilde{p}_{n+1}(z,t)&=&2\int_0^t\D s\int_{z+\sigma_1}^\infty\D y\,\omega(y-z)\sum_{i=0}^{n}\Theta_i(y)\tilde{p}_i(y,s)\label{ptilden1}
\ee
and the line coverage is given by
\be
\phi(t)=1-\int_0^\infty\D z\,z\,\tilde{p}(z,t).\label{tildephi}
\ee
The $t\to\infty$ behaviour of $\tilde{p}_1,...,\tilde{p}_n$ can be derived in analogy to the discussion in Sec.~\ref{Sec:time} showing again an exponential decay governed by $e^{-\psi(z)t}$.

Further below, we need the following expression for the rate of change $\dot{\phi}(t)$. With Eqs.~(\ref{polyms},\ref{tildephi}) we obtain
\be
\dot{\phi}(t)&=&-\int_0^\infty\D z\,z\,\frac{\partial}{\partial t}\tilde{p}(z,t)\nonumber\\
&=&\int_{\sigma_1}^\infty\D z\,z\,\psi(z)\tilde{p}(z,t)-2\int_0^\infty\D z\,z\,\left<\int_{z+\sigma}^\infty\D y\,\tilde{p}(y,t)\right>\nonumber\\
&=&\int_{\sigma_1}^\infty\D z\,z\,\psi(z)\tilde{p}(z,t)-\int_0^\infty\D z\,z^2\,\left<\tilde{p}(z+\sigma,t)\right>\nonumber\\
&=&\int_{\sigma_1}^\infty\D z\,z\,\psi(z)\tilde{p}(z,t)-\left<\int_\sigma^\infty\D z\,(z-\sigma)^2\,\tilde{p}(z,t)\right>\nonumber\\
&=&\int_{\sigma_1}^\infty\D z\left<\sigma(z-\sigma)\Theta(z-\sigma)\right>\tilde{p}(z,t)\nonumber\\
&=&\int_{\sigma_1}^\infty\D z\,\Lambda(z)\sum_{j=0}^n\Theta_j(z)\tilde{p}_j(z,t)
\label{phirate}
\ee
where we define $\Lambda(z)=\left<\sigma(z-\sigma)\Theta(z-\sigma)\right>$ in the last step.

\subsection{Power-law size distribution}

In order to find an optimal size distribution that leads to a maximally dense polydisperse packing, we follow the same approach as in the interacting particle case and consider a power-law size distribution
\be
\label{rhopower}
\rho(\sigma)=\frac{\mu}{\sigma_2-\sigma_1}\left(\frac{\sigma-\sigma_1}{\sigma_2-\sigma_1}\right)^{\mu-1},
\ee
such that the corresponding $\omega(z)$ is identical to the one resulting from the potential Eq.~\eqref{mupot}. However, it is important to note that the resulting dynamics of $\tilde{p}$ is different from that of the interacting particle case for finite $t$. The reason is that $\omega(x)=0$ for $x\le\Delta$, but $\psi(x)=0$ for $x\le 2\Delta$. In the polydisperse case, we have $\omega(z)=0$ and $\psi(z)=0$ for $z\le \sigma_1$. However, the stationary distribution is identical in both cases for $\mu\gg 1$. With Eq.~\eqref{ptilden1} we obtain
\be
\label{pzpoly}
\tilde{p}_{\rm s}(z)&=&\lim_{t\to\infty}\tilde{p}_{n+1}(z,t)=2\Theta(\sigma_1-z)\int_0^\infty\D t\int_{z+\sigma_1}^\infty\D y\,\omega(y-z)\sum_{i=0}^n\tilde{\Theta}_i(y)\tilde{p}_i(y,t).
\ee
For $\mu\gg 1$ we have $\omega(z)\to\Theta(z-\sigma_2)$, $\sigma\to\sigma_2$, $\langle e^{-\sigma t}\rangle\to e^{-\sigma_2t}$ such that
\be
\tilde{p}_{\rm s}(z)&=&2\Theta(\sigma_1-z)\int_0^\infty\D t\int_{z+\sigma_2}^\infty\D y\,\tilde{p}_0(y,t)
\ee
where with Eq.~(\ref{ptilde0})
\be
\int_{z+\sigma_2}^\infty\D y\,\tilde{p}_0(y,t)&=&t\,F(t)e^{-zt},\qquad F(t)\to e^{-2\int_0^t\D s\frac{1-e^{-\sigma_2s}}{s}}.
\ee
Therefore, the stationary distribution of empty gaps is given by
\be
\tilde{p}_{\rm s}(z)&=&\Theta(\sigma_1-z)\tilde{p}^*(z)
\ee
using Eq.~\eqref{ptildestat} with $\sigma_2=a$. With Eq.~\eqref{tildephi} we obtain the jamming density
\be
\phi_{\rm J}&=&1-\int_0^{\sigma_1}\D z\,z\,\tilde{p}_{\rm s}(z)\nonumber\\
&=&1-\int_0^{\sigma_2}\D z\,z\,\tilde{p}_{\rm s}(z)+\int_{\sigma_1}^{\sigma_2}\D z\,z\,\tilde{p}_{\rm s}(z)\nonumber\\
&=&\phi_{\rm R}+\int_{\sigma_1}^{\sigma_2}\D z\,z\,\tilde{p}_{\rm s}(z),
\ee
which agrees with Eq.~\eqref{maxphi} provided we also set $\sigma_1=2\Delta-a$.

\subsection{The case $n=1$}

For the special case $\sigma_1<\sigma_2\le 2\sigma_1$ a result for the jamming density $\phi_{\rm J}$ has been obtained in Ref.~\cite{Burridge:2004aa} using a recursive approach. We reproduce this result in the following. With Eq.~\eqref{phirate} we have the line coverage as a function of time
\be
\phi(t)&=&\int_{\sigma_2}^\infty\D z \,(z\overline{\sigma}-\left<\sigma^2\right>)\int_0^t\D s\,\tilde{p}_0(z,s)+\int_{\sigma_1}^{\sigma_2}\D z\,\Lambda(z)\int_0^t\D s\,\tilde{p}_1(z,s)\nonumber\\
&=&\int_0^t\D s\,F(s)e^{-(\sigma_2-\overline{\sigma})s}\left(\overline{\sigma}(1+\sigma_2s)-\left<\sigma^2\right>s\right)+\int_{\sigma_1}^{\sigma_2}\D z\,\Lambda(z)\int_0^t\D s\,\tilde{p}_1(z,s)
\label{phiraten1}
\ee
calculating the $z$-integral in the first term. The distribution $\tilde{p}_1$ is given by Eq.~\eqref{pjpoly}. Using Eq.~\eqref{omegapoly} we obtain
\be
\tilde{p}_1(z,t)&=&2\int_0^t\D s\,e^{-\psi(z)(t-s)}\left<\int_{z+\sigma}^\infty\D y\,\tilde{p}_0(y,s)\right>\nonumber\\
&=&2\int_0^t\D s\,e^{-\psi(z)(t-s)}\tilde{p}_0(z,s)\frac{\left<e^{-\sigma s}\right>}{s}\nonumber\\
&=&2\int_0^t\D s\,e^{-\psi(z)(t-s)-zs}\,s\,F(s)\left<e^{-(\sigma-\overline{\sigma})s}\right>.
\ee
Substituting $\tilde{p}_1$ in Eq.~\eqref{phiraten1} and taking the $t\to\infty$ limit yields
\be
\phi_{\rm J}=\int_0^\infty\D s\,F(s)e^{-(\sigma_2-\overline{\sigma})s}\left(\overline{\sigma}(1+\sigma_2s)-\left<\sigma^2\right>s\right)+\int_0^\infty\D s\,s\,F(s)\left<e^{-(\sigma-\overline{\sigma})s}\right>\int_{\sigma_1}^{\sigma_2}\D z\,\frac{\Lambda(z)}{\psi(z)}e^{-zs}
\ee
after a partial integration, which is the expression obtained in \cite{Burridge:2004aa}.

\subsection{Asymptotic scaling}

Since the $\tilde{p}_j$ for $j=0,...,n-1$ all decay exponentially as $t\to\infty$, the asymptotic approach to $\phi_{\rm J}$ is governed by $\tilde{p}_n(z,t)\simeq B_n(z)e^{-\psi(z)t}$ in analogy to Sec.~\ref{Sec:time}. The scaling of $\phi_{\rm J}-\phi(t)$ could thus be calculated as in Sec.~\ref{Sec:asymp2}, but here we use instead Eq.~\eqref{phirate} to obtain
\be
\phi_{\rm J}-\phi(t)&\simeq& \int_t^\infty\D s\int_{\sigma_1}^{\sigma_2-(n-1)\sigma_1}\D z\,\Lambda(z)\tilde{p}_n(z,s)\nonumber\\
&\simeq&\int_{\sigma_1}^{\sigma_2-(n-1)\sigma_1}\D z\,\frac{\Lambda(z)}{\psi(z)}B_n(z)e^{-\psi(z)t}.
\ee
As $z\to\sigma_1$, the minimum of $\psi(z)$, we have $\Lambda(z),\psi(z)\to 0$, while $B_n(z)$ remains finite (e.g., for $n=1$, $B_n=\tilde{p}_1(z,t_{\rm c})e^{\psi(z)t_{\rm c}}$). The asymptotic scaling thus depends on the behaviour of $\Lambda(z)/\psi(z)$ as $z\to\sigma_1$. We obtain
\be
\Lambda'(z)&=&\int_{\sigma_1}^z\D u\,u\,\rho(u), \qquad\qquad\psi'(z)=\int_{\sigma_1}^z\D u\,\rho(u)\\
\Lambda''(z)&=&z\rho(z), \qquad\qquad \psi''(z)=\rho(z)
\ee
and thus
\be
\lim_{z\to\sigma_1}\frac{\Lambda(z)}{\psi(z)}=\lim_{z\to\sigma_1}\frac{\Lambda''(z)}{\psi''(z)}=\lim_{z\to\sigma_1}z=\sigma_1.
\ee
Therefore
\be
\phi_{\rm J}-\phi(t)&\sim&\int_{\sigma_1}^{\sigma_2-(n-1)\sigma_1}\D z\,e^{-\psi(z)t}\sim t^{-1/\alpha},
\ee
assuming the asymptotic form
\be
\psi(z)\sim (z-\sigma_1)^\alpha,\qquad z\to\sigma_1.
\ee
Since $\psi(\sigma_1)=\psi'(\sigma_1)=0$ and $\psi''(z)=\rho(z)$, the exponent $\alpha$ is determined by the leading term in the asymptotic expansion of $\rho$ as $z\to\sigma_1$. For the general power-law size distribution Eq.~\eqref{rhopower} we obtain
\be
\psi(z)\sim (z-\sigma_1)^{\mu+1},\qquad\qquad\phi_{\rm J}-\phi(t)\sim t^{-1/(\mu+1)}.
\ee

\end{appendix}

\end{document}